%% file: main.tex
\documentclass[conference]{IEEEtran}

\usepackage{enumitem}    

\usepackage[table,dvipsnames]{xcolor}
\usepackage{enumitem}    
\newif\ifoutline
\outlinetrue

\usepackage{tikz}
\newcommand*\circled[1]{\tikz[baseline=(char.base)]{ \node[shape=circle,fill=black,inner sep=0.4pt] (char) {\textcolor{white}{#1}};}}

\usepackage{amsmath}
\usepackage{graphicx}
\usepackage{xcolor}
\usepackage{multirow}
\usepackage{multicol}
\usepackage{booktabs}
\ifCLASSINFOpdf
\else
\fi
\renewcommand{\IEEEbibitemsep}{0pt plus 0.5pt}
\makeatletter
\IEEEtriggercmd{\reset@font\normalfont\fontsize{7.9pt}{8.60pt}\selectfont}
\makeatother
\IEEEtriggeratref{1}

\makeatletter
\newcommand*\titleheader[1]{\gdef\@titleheader{#1}}
\AtBeginDocument{%
  \let\st@red@title\@title
  \def\@title{%
    \bgroup\normalfont\normalsize\centering\@titleheader\par\egroup
    \vskip1ex\st@red@title}
}
\makeatother

\titleheader{\vspace{-30pt}\color{gray}To appear at the IEEE International Symposium on Circuits and Systems (ISCAS'25), May 25-28, London, United Kingdom.}
\title{A Bespoke Design Approach to Low-Power Printed Microprocessors for Machine Learning Applications}

\begin{document}

\author{
    \IEEEauthorblockN{
        Panagiotis Chaidos,
        Giorgos Armeniakos,
        Sotirios Xydis
        and Dimitrios Soudris
    }
    \IEEEauthorblockA{
        National Technical University of Athens, Greece\\ \{pchaidos, armeniakos, sxydis, dsoudris\}@microlab.ntua.gr,
    }
}

\maketitle

\begin{abstract}
Printed electronics have gained significant traction in recent years, presenting a viable path to integrating computing into everyday items, from disposable products to low-cost healthcare.
However, the adoption of computing in these domains is hindered by strict area and power constraints, limiting the effectiveness of general-purpose microprocessors.
This paper proposes a bespoke microprocessor design approach to address these challenges, by tailoring the design to specific applications and eliminating unnecessary logic. Targeting machine learning applications, we further optimize core operations by integrating a SIMD MAC unit supporting 4 precision configurations that boost the efficiency of microprocessors.
Our evaluation across 6 ML models and the large-scale Zero-Riscy core, shows that our methodology can achieve improvements of 22.2\%, 23.6\%, and 33.79\% in area, power, and speed, respectively, without compromising accuracy.
Against state-of-the-art printed processors, our approach can still offer significant speedups, but along with some accuracy degradation. This work explores how such trade-offs can enable low-power printed microprocessors for diverse ML applications.
\end{abstract}


%

\input{introduction}
\input{related}

\input{bespoke_optimizations}

\input{evaluation}

\input{conclusion}

\section*{Acknowledgment}
This work is funded in part by the Convolve project evaluated by the
EU Horizon Europe research and innovation program under grant agreement
No. 101070374

{
\bibliographystyle{IEEEtran}
\bibliography{references}
}


\end{document}

%% file: introduction.tex
\section{introduction}

In recent years, printed electronics have emerged as a key player in the Internet of Things (IoT) landscape, particularly for ultra-resource constrained devices~\cite{Bleier:ISCA:2020:printedmicro}. 
These technologies are central to embedded machine learning (ML) inference, offering mechanical flexibility, low non-recurring engineering (NRE) costs, and affordability in production. 
As a crucial component of the "Fourth Industrial Revolution," printed electronics enable the development of smart, lightweight devices using inexpensive, widely available materials and simple printing processes, paving the way for innovative integration methods and functionality at reduced costs~\cite{hobbie2024conformal}.

Printed electronics present an opportunity to bring computing and intelligence into sectors like disposable products (e.g., packaged foods and beverages), smart packaging, low-cost healthcare (e.g., smart bandages), in-situ monitoring, and the expansive fast-moving consumer goods (FMCG) market~\cite{Kumar2017Bespoke}. 
However, these domains have seen limited adoption due to strict area and power constraints, particularly in wearables and implantables. 
While high costs drive applications to use general-purpose microprocessors and microcontrollers, their inefficiency in meeting the demands of these applications highlights a growing need for custom (\textit{bespoke}) microprocessors.

Bespoke~\cite{Kumar2017Bespoke,Ozer2019Bespoke} microprocessors enhance area and power efficiency by removing unused logic specific to a given application. 
By removing unnecessary components, these custom designs—tailored to one or several applications, significantly reduce both area and power consumption compared to general-purpose microprocessors.
Removing unnecessary logic enables further optimization through the modification (or even addition~\cite{armeniakos2024mixed}) of hardware units, tailoring them to the application's specific operations. 
As a result, performance can be significantly improved, as well.

In this work, we embrace the bespoke design paradigm and apply a set of bespoke logic reductions to boost area and power efficiency.
As a proof-of-concept scenario we profile a set of 2 low power cores, i.e., Zero-Riscy of PULP platform and TP-ISA~\cite{Bleier:ISCA:2020:printedmicro}\footnote{We use TP-ISA and Zero-Riscy as our proof-of-concept microarchitecture. The proposed workflow can be straightforwardly adopted in other processors.}.
Continuously, by leveraging the freed-up area and focusing on core operations for ML applications, we modify existing ALU and develop an SIMD MAC unit with support for multiple precision levels.
Overall, compared to our baseline zero-riscy core and across 3 MLPs and 3 SVMs models, our approach achieves 22.2\%, 23.6\% and 33.79\% improvements in area, power and speedup, respectively, with zero accuracy loss.
On the other hand, when compared to the state-of-the-art printed processor TP-ISA, our methodology delivers an $85.1\%$ speedup, albeit with trade-offs of 1.98x area, 1.82x power, and a 0.5\% top-1 accuracy loss.

\textbf{Our main contributions in this work are the following:}

\begin{enumerate}[topsep=0pt,leftmargin=*]
    \item We propose a generic methodology for bespoke microprocessors - an approach to reducing area and power by tailoring a processor to specific applications.
    \item Leveraging our bespoke design paradigm, we further optimize the ALU unit of the examined processors by incorporating an SIMD MAC unit for several precision levels.
    \item This is the first work to synthesize large-scale microprocessors for printed technologies. Despite the trade-offs explored, our findings represent a significant step toward enabling battery-powered printed microprocessors.
\end{enumerate}

%% file: related.tex
\section{Related Work \& Background}


Printed electronics cannot match silicon-based electronics in terms of integration density, area, or performance. Typical operating frequencies for printed circuits range from a few Hz to a few kHz~\cite{cadilha2017digital}, and feature sizes are usually several microns~\cite{lei2019low}. Despite these limitations, printed electronics offer unique advantages such as flexibility, adaptability, and most notably, drastically lower fabrication costs — even at low volumes — making them ideal for application domains that traditional silicon-based VLSI cannot easily access.

The field of printed electronics has seen a substantial increase in research across multiple application areas. One example is the development of RFID tags utilizing pseudo-CMOS logic to enhance thin-film circuit performance~\cite{Myny:2021:dualinput}. Another notable effort involved fabricating a 2-input neuron designed to perform MAC operations~\cite{Weller:ASPDAC:2020}. More recently, ARM achieved a breakthrough by creating a flexible 32-bit microprocessor comprising over $18,000$ gates~\cite{Biggs:Nature2021:flexarm}
.

While several studies focus on optimization and approximation techniques to mitigate area and power limitations of printed circuits~\cite{arm2023codesign,arm2023crossapprox}, research on printed ML applications remains in its early stages, mainly due to the large feature sizes of printed circuits. For instance, \cite{Ozer2019Bespoke} introduced an automated approach for creating bespoke classifiers, and in~\cite{Ozer:Nature:2020}, a hardwired machine learning processor was integrated into a system for odor recognition. Additionally, \cite{Weller:2021:printed_stoch}
 explored Stochastic Computing (SC) to reduce the area and power of printed MLPs, but this often resulted in significant accuracy loss.
This work goes along with state-of-the-art, investigates the potential of printed microprocessors using printed technology and evaluates how bespoke, optimized synthesis methods can facilitate the creation of efficient printed cores.

%% file: bespoke_optimizations.tex
\section{Enabling Bespoke Microprocessors}

This section outlines the workflow for extracting hardware specifications, including ROM usage, processor area, timing, and power, as well as the process of eliminating underutilized hardware. We present a methodology for measuring hardware utilization according to the specific needs of the applications. The non-utilized hardware blocks and architectural components are removed based on information gathered from the workflow. Finally, we develop an SIMD Multiply-Accumulate(MAC) unit to accelerate the operations of ML models, utilizing several levels of precision to enable parallel computation and achieving  significant speedup. The proposed methodology is demonstrated through a proof-of-concept implementation on two low-power processors.


  \subsection{Application Dependent Logic Reduction}

This work investigates a suite of ML applications designed for printed computing, utilizing the EGFET standard cell library. Given the flexibility advantage of microprocessors compared to Application Specific Integrated Circuits(ASICs), the study profiles and applies the proposed methodology on two low-power microprocessors, Zero-Riscy, a 32-bit 2-stage pipeline RISC-V architecture, and TP-ISA~\cite{Bleier:ISCA:2020:printedmicro}, a minimal highly configurable core, as a proof-of-concept.

\begin{figure}[t]
   \centering
  \includegraphics[width=0.35\textwidth]{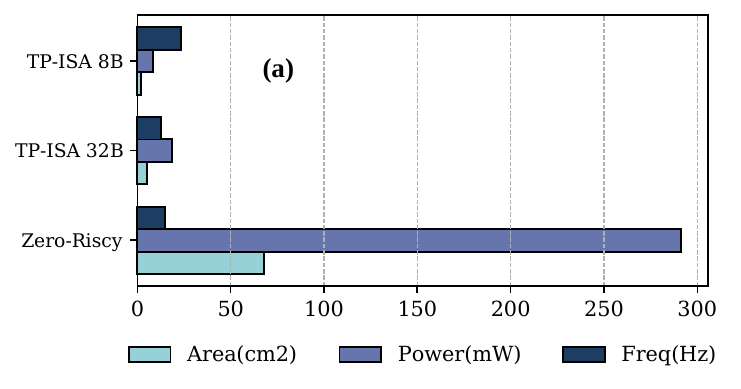}\\
  \includegraphics[width=0.35\textwidth]{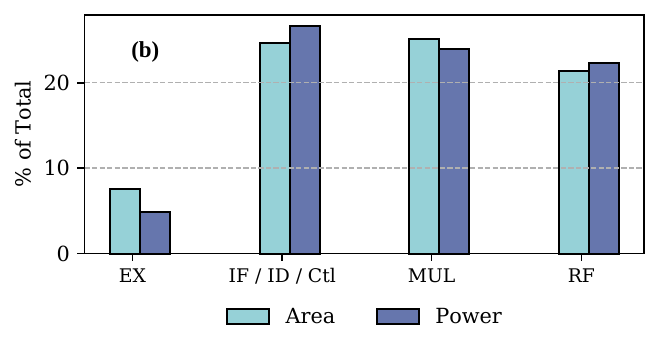}
  \caption{\textbf{a}): Baseline Area, Power and System Clock for Zero-Riscy and TP-ISA for EGFET printed technology. \textbf{b}): Percentage of total area and power consumption for the main functional units of Zero-Riscy, \textbf{EX}(Execution Unit), \textbf{MUL}(Multiplier), \textbf{RF}(Register File) and \textbf{IF / ID / Ctl}(Instruction Fetch, Instruction Decode and Controller units) grouped together. }
  \label{fig:base_cores}
\end{figure}

We synthesize Zero-Riscy and two configurations of TP-ISA. As shown in Figure \ref{fig:base_cores}, the total area and power consumption of both cores are presented, along with the percentage data for the large functional units of Zero-Riscy, since TP-ISA does not have clearly defined components. Zero-Riscy, being a larger-scale low power processor, appears to be  prohibitively large in the EGFET technology, reaching a circuit area of 67.53 cm2 and power consumption of 291.21mW. It is worth noting that the multi-stage multiplier unit and the register file of Zero-Riscy account for almost half of the total area and power consumption, at 46.5\% and 46.2\% respectively. In contrast, both TP-ISA configurations fall well within the technology limitations, suggesting a focus on performance optimizations rather than area and power consumption. Furthermore, when considering printed memories, code length can also significantly reduce area and power consumption by utilizing fewer ROM cells. Each ROM cell takes up 0.84mm2 and 18.23uW, favoring designs with narrower bit-widths and smaller code sizes.

We assemble a collection of printed ML and associated applications~\cite{zhai2009energy,Bleier:ISCA:2020:printedmicro}, including a 3-layer Multi-Layer Perceptron(MLP), a depth-2 Decision Tree(DT), simple Multiplication-Division and Insertion Sort on array of size 16. For Zero-Riscy, the Debug, Interrupt Controller, and Compressed Decoder Unit are not utilized and are completely removed. From the RISC-V instruction set, the SLT, most CSR, System Calls, and MULH instructions remain unused and can be effectively eliminated. Additionally, 12 registers are sufficient for executing all benchmarks, allowing for the removal of the rest. Since the code size and register usage for most applications are less than the baseline, this means that they can be addressed with narrower bitwidths, enabling a reduction in the Program Counter(PC) from 32 bits to 10 bits, and a reduction in the Base Address Registers(BARs) from 32 bits to 8 bits. TP-ISA is proven to be minimal and thus the focus here is primarily on improving performance and reducing code size, as will be discussed next.



 \subsection{ML acceleration unit}

 Logic reductions executed on the previous step create space for extensions that should lead to performance improvements. In order to optimize for ML workloads, we design a unit targeting the most frequent and computationally intensive ML operation - MAC. Most ML models heavily utilize MAC, which is mainly used for computing neuron activations. The proposed unit enables single-cycle multiplication and accumulation, significantly improving performance compared to the baseline, which require at least 3 cycles for Zero-Riscy and several more for TP-ISA where the whole operation is scheduled to the ALU. Additionally, the reduced instruction count for MAC-intensive code leads to program memory savings through smaller ROM requirements.

\begin{figure}[t]
  \centering
  \includegraphics[width=0.45\textwidth]{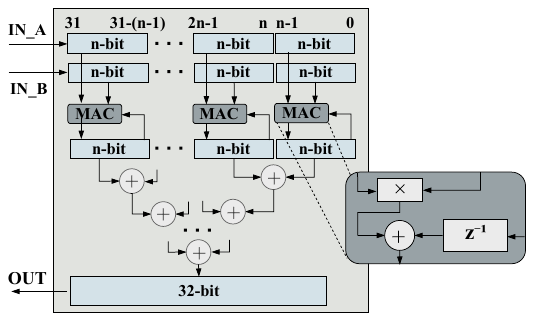}
  \caption{Overview of proposed MAC unit. The unit has been implemented for precision options with \(n\)=32, 16, 8 and 4 bits. For each option, the unit can be split into 1, 2, 4 and 8 concurrent operations respectively.}
  \label{fig:MAC_Unit}\vspace{-2ex}
\end{figure}

ML models exhibit inherent resilience to approximation errors, maintaining relatively high model accuracy up to certain fault thresholds~\cite{nn_tolerance}. We leverage the tolerance of these applications by integrating several precision options for the proposed MAC units. This design choice permits us to (a)utilize parallel execution  and (b)replace large multipliers with small ones that have less depth. Figure~\ref{fig:MAC_Unit} depicts the architecture of the unit for a selected precision \(n\). By employing precision \(n\), the unit effectively computes MAC of \(32/n\) neurons in a single cycle, as shown in Equation~\ref{eq:MAC_Unit}.

\begin{equation}
    \footnotesize
    \begin{aligned}
        &acc_{total} = \sum_{i=1}^{K}{acc_i} \\      
        &acc_1 = (r1[n-1:0] \times r2[n-1:0]) + acc_1 \\
        &acc_2 = (r1[2n-1:n] \times r2[2n-1:n]) + acc_2 \\
        &\vdots  \\
        &acc_k = (r1[31:31-(n-1)] \times r2[31:31-(n-1)]) + acc_k \\  
    \label{eq:MAC_Unit}
    \end{aligned}
\end{equation}


 \subsection{Workflow}

\begin{figure}[t]
  \centering
  \includegraphics[width=0.48\textwidth]{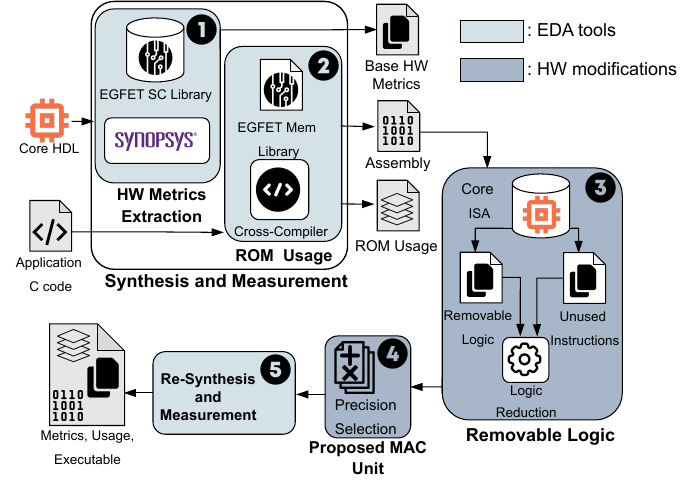}
  \caption{Diagram of the proposed methodology for bespoke ML microprocessors.}
  \label{fig:methodology}\vspace{-2ex}
\end{figure}
 
The complete overview of our proposed workflow for generating bespoke microprocessors is presented in Figure~\ref{fig:methodology}.
Firstly, using Synopsys DC with the EGFET library we synthesize the cores and extract the area and power consumption information of the design(\circled{1}). With each benchmark's code, using the respective compilers, the executables and assembly code are produced. Using the size of the executables along with the overhead of ROM cells in EGFET, we calculate the total area and power consumption of the program memory required to store each benchmark(\circled{2}). 

Utilizing the compiled assembly code, we extract unused instructions for each ISA and remove all unused HW components and unused logic. We analyse the use of architectural components including flags and registers. By looking at the code sizes and utilized registers we determine the required bitwidth for addressing, thus trimming the PC and BAR(\circled{3}). 

Having eliminated unused logic, we look towards performance optimizations and insert our proposed SIMD MAC unit as part of the ISA and test with several precision configurations(\circled{4}). The benchmarks are rewritten to be executed on the unit, reducing code length and optimizing program memory. Finally, the RTL is passed in the HW measurement process again to obtain the area and power information(\circled{5}).

%% file: evaluation.tex
\section{Results \& Analysis}

\subsection{Experimental Setup}

To evaluate our proposed methodology, we assess the performance of four models: MLP-C, MLP-R, SVM-C, and SVM-R, which are trained on the Cardiography, RedWine, and WhiteWine datasets from the UCI repository~\cite{uci}. The models are trained using the scikit-learn library, employing randomized parameter optimization (RandomizedSearchCV) with a 5-fold cross-validation procedure. Input features are normalized to the range [0, 1], and the data is split into training and test sets with a 70\%/30\% ratio. For the MLP models, a single hidden layer with up to five neurons is used, and the ReLU activation function is applied. The SVM models use a linear kernel, with SVM-C models implementing a one-vs-one classification strategy. The architecture of each MLP is configured to use the minimal number of hidden nodes while ensuring that all MLPs achieve near-max accuracy.

For RTL simulation of the two cores that underwent our proposed methodology, we use Modelsim with the compiled executable for each model. After applying the optimizations of our approach, we implement MAC units with various levels of precision for each core. The smallest 4-bit TP-ISA is realized with a 4-bit MAC unit and no parallelization, as the bitwidth is insufficient to support it. The 32-bit TP-ISA and Zero-Riscy were assessed with MACs of 4-bit, 8-bit, 16-bit, and 32-bit precision, with the maximum parallelization allowed for 16 bits or less, as shown in Figure~\ref{fig:MAC_Unit}.

\subsection{Evaluation}



\begin{table}[t]
    \centering
    \caption{Percentage area-power gains, average speedup and error of bespoke Zero-Riscy(ZR) Core.\textbf{B} is Bespoke and \textbf{P} is Precision. Precision implementations utilize parallelization up to 32 bits}
    \label{tab:pulpino_results}
    \begin{tabular}{|l|c|c|c|c|}
        \hline
        \textbf{Cores} & \textbf{Area} & \textbf{Power} & \textbf{Speedup} & \textbf{Accuracy Loss} \\
        \hline
        \textbf{ZR B} & 10.6\% & 11.4\% & 0\% & 0.0\% \\
        \textbf{ZR B MAC 32} & 8.2\% & 14.4\% & 23.93\% & 0.0\% \\
        \textbf{ZR B MAC P16} & 22.2\% & 23.6\% & 33.79\% & 0.0\% \\
        \textbf{ZR B MAC P8} & \textbf{29.3}\% & \textbf{28.7}\% & \textbf{41.73}\% & \textbf{0.5}\% \\
        \textbf{ZR B MAC P4} & \textbf{36.5\%} & \textbf{34.1\%} & \textbf{46.4\%} & \textbf{15.66\%} \\
        \hline
    \end{tabular}
\end{table}

\begin{figure}[t]
  \centering
  \includegraphics[width=0.25\textwidth]{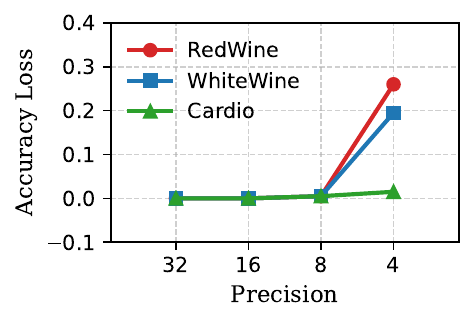}
  \caption{Average Accuracy Loss per Model introduced by each Precision Option}
  \label{fig:model_error}
\end{figure}

Table~\ref{tab:pulpino_results} shows the exact gains and error trade-offs of our methodology, for several levels of precision, on the proof-of-concept Zero-Riscy compared to the baseline. Since all the models' parameters are 16-bits, we observe that we can parallelize the unit with 16 bit accuracy, gaining in all fronts and sacrificing no accuracy. In more detail, Figure~\ref{fig:model_error} shows no error from 32 down to 16 bits, a small increase for 8 bits and a jump for most models at 4 bits, reaching a prohibitive 26\% for RedWine. Balancing the trade-offs, 8-bit precision appears to be a suitable compromise, introducing just 0.5\% average decrease in accuracy.    

\begin{figure}[t]
  \centering
  \includegraphics[width=0.5\textwidth]{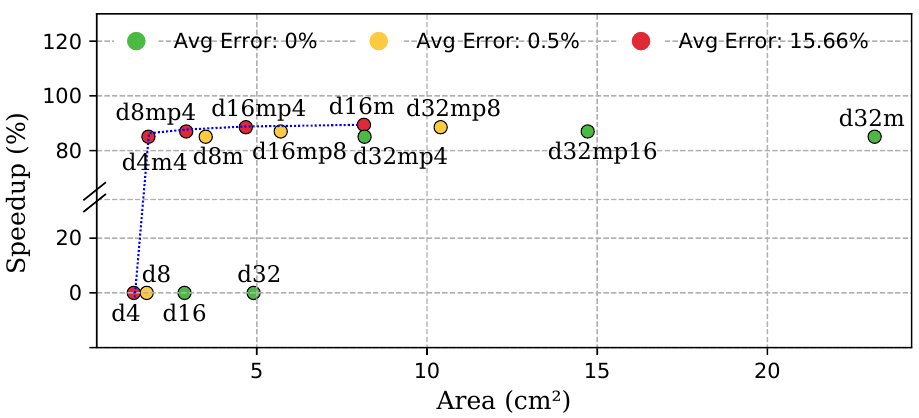}
  \caption{Scatterplot of TP-ISA configurations, where \textbf{d} is the bits of the datapath, \textbf{m} signifies that the proposed MAC unit is implemented (with d bits) and \textbf{p} is the precision of the unit (lack of p means that the precision is the standard of the core and so there is no parallelization). The Pareto Front for Area and Speedup is highlighted in blue.}
  \label{fig:scatterplot}\vspace{-2ex}
\end{figure}

\begin{table}[t]
  \centering
  \caption{Gains, average speedup and error of bespoke 8-bit TP-ISA. Pareto solution, maintaining relatively low average error increase}
  \label{tab:8-bit_tp_isa}
  \begin{tabular}{|c|c|c|}
    \hline
    \textbf{Configuration} & \textbf{TP-ISA 8-BIT MAC} \\
    \hline
    Area Overhead & x1.98\\
    \hline
    Power Overhead & x1.82\\
    \hline
    Avg Err (Base is 0.5\%) & 0.5\%\\
    \hline
    Estimated Speedup & \textbf{up to 85.1\%} \\
    \hline
  \end{tabular}
\end{table}

Figure~\ref{fig:scatterplot} shows all base and generated TP-ISA~\cite{Bleier:ISCA:2020:printedmicro} designs with the blue line highlighting the Area-Speedup Pareto curve. This curve remains similar even when considering power, as area and power exhibit a near-linear correlation in these examples. The lower-left group of points corresponds to the baseline cores, achieving no speedup, while the upper-side implementations are generated through the proposed methodology. Speedup increases rapidly when using a MAC unit and then slowly with SIMD. Table~\ref{tab:8-bit_tp_isa} shows a Pareto solution that achieves substantial speedup with minimal accuracy degradation of 0.5\% and overhead factors of 1.98x and 1.82x for area and power, remaining still well within printed batteries' capabilities while vastly speeding up execution time by 85.1\% compared to~\cite{Bleier:ISCA:2020:printedmicro}.

When considering printed memories, we observe that (a)architectures with smaller bitwidths benefit more in terms of memory overheads due to direct reduction of cells per addressable space, (b)architectures that support multiplication save up to 11.1\% on memory, as the  multiplication instructions do not need to be scheduled for ALU being directly replaced with a single MUL command and (c)configurations that employ SIMD can introduce additional savings of up to 1-2\% by calculating entire neurons in a single pass, without requiring additional control instructions for loops.

%% file: conclusion.tex
\section{conclusion}

Printed electronics emerge as a promising technology, enabling low-cost, lightweight, battery powered applications, especially when paired with the flexibility of general purpose processors. However, the advantages of printed processors come at the cost of prohibitively large, power hungry and lower performance circuits. Our work explores the trade-offs of the printed computing design space and demonstrates that bespoke optimizations on low power processors, targeting ML applications, achieves consistent improvements in performance and hardware characteristics while offering a series of Pareto-optimal solutions considering trade-offs of area, power, speedup and accuracy loss for printed electronics use cases.